\documentclass[prd,aps,floatfix,nofootinbib,11 pt]{revtex4}
\usepackage{amssymb}
\usepackage{amsmath,graphicx,color,epsfig}

\input{tcilatex}

\begin{document}

\title{Environment-assisted quantum Minority games}
\author{M. Ramzan\thanks{%
mramzan@phys.qau.edu.pk} and M. K. Khan}
\address{Department of Physics Quaid-i-Azam University \\
Islamabad 45320, Pakistan}

\date{\today}

\begin{abstract}
The effect of entanglement and correlated noise in a four-player quantum
Minority game is investigated. Different time correlated quantum memory
channels are considered to analyze the Nash equilibrium payoff of the 1st
player. It is seen that the Nash equilibrium payoff is substantially
enhanced due to the presence of correlated noise. The behaviour of damping
channels (amplitude damping and phase damping) is approximately similar.
However, bit-phase flip channel heavily influences the minority game as
compared to other channels in the presence of correlated noise. On the other
hand, phase flip channel has a symmetrical behaviour around $50\%$ noise
threshold. The significant reduction in payoffs due to decoherence is well
compensated due to the presence of correlated noise. However, the Nash
equilibrium of the game does not change in the presence of noise. It is seen
that in case of generalized amplitude damping channel, entanglement plays a
significant role at lower level of decoherence. The channel has less
dominant effects on the payoff at higher values of decoherence. Furthermore,
amplitude damping and generalized amplitude damping channels have almost
comparable effects at lower level of decoherence $(p<0.5)$. Therefore, the
game deserves careful study during its implementation due to prominent role
of noise for different channels.
\end{abstract}

 \maketitle


\vspace*{1.0cm}

\vspace*{1.0cm}

%\newpage

\section{Introduction}

In the recent past, rapid interest has been developed in the discipline of
quantum information \cite{NieMA} that has led to the creation of quantum
game theory \cite{MeyDA}. During last few years, number of authors have
contributed to the development of quantum game theory [3-14]. Several
classical games have been converted into quantum domain such as quantum
prisoners' dilemma [15-31]. James et al. [32] have analyzed the quantum
penny flip game using geometric algebra. Almeida et al. [33] have suggested
that quantum correlations provide no advantage over classical correlations
in a multipartite nonlocal game. Recently, Sharif and Heydari [34] have
investigated Minority games for various initial states with different level
of entanglement. They have shown that with the aid of entanglement and
linear superposition of strategies, quantum games are shown to yield
significant advantage over their classical counterparts. Some more recent
investigations in the field of quantum game theory include the contributions
from different authors, for details, see references [35-41].

Noise effects in different quantum games have been investigated by many
authors [5, 7, 10, 25] and found interesting results. In Ref. [42], we have
studied noise effects in quantum magic squares game. It is shown that the
probability of success can be used to determine the characteristics of
quantum channels. Implementation of decoherence and correlated noise has
also been extended to the novel field of quantum information theory [43,
44]. The Minority game has received much attention as a model of a
population of agents repeatedly buying and selling in a market [45, 46].
First quantum version of a four player quantum Minority game (QMG) was
examined by Benjamin and Hayden [47], and later generalized to $N$-players
[48]. Flitney et al. have extended its consideration towards the
implementation of decoherence [49]. Quantum channels with memory [50-52]
provides a natural theoretical framework for the study of any noisy quantum
communication system where correlation time is longer than the time between
consecutive uses of the channel. A more general model of a quantum channel
with memory was introduced by Bowen and Mancini [53] and also studied by
Kretschmann and Werner [54].

In this paper, we analyze a four-player quantum Minority game influenced by
different time correlated quantum memory channels, such as amplitude
damping, depolarizing, bit-phase-flip and phase flip channels, parameterized
by decoherence parameter $p$\ and memory parameter $%
%TCIMACRO{\U{3bc} }%
%BeginExpansion
\mu
%EndExpansion
$. Here $p\in \lbrack 0,1]$ and $\mu \in \lbrack 0,1]$ represent the lower
and upper limits of decoherence parameter and memory parameter respectively.
In addition, the generalized amplitude-damping channel, the most prominent
representative of non-unital channels, is also considered parameterized by
the decoherence parameter $p$ and parameter $\alpha \in \lbrack 0,1]$ that
depends on the temperature of the environment. It is seen that the players
payoffs heavily depends on the memory of the channel under consideration. A
similar behaviour of depolarizing and bit-phase flip channels is seen for
maximum correlations. Amplitude damping channel influences the game more
heavily as compared to the other channels. Whereas, the generalized
amplitude damping channel reveals lower dissipation effects when compared
with amplitude damping channel at higher level of decoherence. The Nash
equilibrium payoff is substantially enhanced due to the presence of quantum
memory. It is shown that memory controls the payoff reduction due to
decoherence.

\section{Quantum Minority games in a correlated environment}

Since noise is a major hurdle in the path of efficient information
transmission from one party to the other. This noise causes a distortion of
the information sent through the channel. Information transmission is said
to be reliable if the probability of error, in decoding the output of the
channel, vanishes asymptotically in several uses of the channel. Let the
initial state of the game consists of one qubit for each player, prepared in
an entangled three-qubit $GHZ$ state by an entangling operator $\hat{J}$
acting on $|0000\rangle $. In the Eisert protocol, this is achieved by
applying\ $\hat{J}^{\dagger }$\ to the game state and then making a
measurement in the computational basis state. Pure quantum strategies are
local unitary operators acting on a player's qubit. In a standard quantum
game protocol, after the execution of players moves, the game state
undergoes a positive operator valued measurement and the payoffs are
determined from the classical payoff matrix, usually given for bi-matrix
games. The classical pure strategies in case of Minority game are always to
choose $0$\textquotedblright\ or $1$. Here we use the methodolgy and
notations of Ref. [49] with additional parameters $\mu $ and $\alpha $ as
defined below in this section. A four-player quantum Minority game in the
presence of time correlated quantum memory channels can be \ described using
the Eisert scheme as%
\begin{eqnarray}
\rho _{0} &=&|\Psi _{0}\rangle \left\langle \Psi _{0}\right\vert \qquad
\qquad \qquad \text{(initial state)} \\
\rho _{1} &=&\hat{J}\rho _{0}\hat{J}^{\dagger }\qquad \qquad \qquad \qquad
\text{(entanglement)} \\
\rho _{2} &=&D(\rho _{1},p_{1},\mu _{1})\qquad \qquad \quad \text{(partial
decoherence and correlations)} \\
\rho _{3} &=&\underset{k_{=1}}{\otimes ^{4}}\hat{M}_{k}\rho _{2}(\underset{%
k_{=1}}{\otimes ^{4}}\hat{M}_{k})^{\dagger }\quad \text{(players' moves)} \\
\rho _{4} &=&D(\rho _{3},p_{2},\mu _{2})\qquad \qquad \text{(partial
decoherence and correlations)} \\
\rho _{5} &=&\hat{J}^{\dagger }\rho _{4}\hat{J}\qquad \qquad \qquad \quad
\text{(preparation for measurement)}
\end{eqnarray}%
to produce the final state $\rho _{f}\equiv \rho _{5}$ upon which a
measurement is taken. Here $\hat{M}_{k}$\ represents the $k^{\text{th}}$
player move. However, for the sake of simplicity, $p_{1}=p_{2}=p$ and $\mu
_{1}=\mu _{2}=\mu $ are used in rest of the calculations. The function $%
D(\rho ,p,\mu )$ represents a completely positive map which can be
completely described in Kraus operator formalism as studied by Macchiavello
and Palma [39] a Pauli channel with partial memory. The two qubit Kraus
operators for such a channel can be written as%
\begin{equation}
A_{ij}=\sqrt{p_{i}[(1-\mu )p_{j}+\mu \delta _{ij}]}\sigma _{i}\otimes \sigma
_{j}
\end{equation}%
where $\sigma _{i}$ ($\sigma _{j})$ are usual Pauli matrices, $p_{i}$ ($%
p_{j} $) represent the quantum noise and indices $i$ and $j$ runs from $0$
to $3.$ The above expression means that with probability $\mu $ the channel
acts on the second qubit with the same error operator as on the first qubit,
and with probability $(1-\mu ),$ it acts on the second qubit independently.
Physically the parameter $%
%TCIMACRO{\U{3bc} }%
%BeginExpansion
\mu
%EndExpansion
$ is determined by the relaxation time of the channel when a qubit passes
through it. In order to remove correlations, one can wait until the channel
has relaxed to its original state before sending the next qubit, however
this lowers the rate of information transfer. The action of a Pauli channel
with memory on $n$-qubits can be generalized in Kraus operator form as%
\begin{equation}
A_{i_{1}.....i_{n}}=\sqrt{p_{i_{n}}\prod\limits_{m=1}^{n-1}[(1-\mu
)p_{i_{m}}+\mu \delta _{i_{m},i_{m+1}}]}\sigma _{i_{1}}\otimes .....\otimes
\sigma _{i_{n}}  \label{KO}
\end{equation}%
\ As stated above, with probability ($1-%
%TCIMACRO{\U{3bc} }%
%BeginExpansion
\mu
%EndExpansion
)$ the noise is uncorrelated and can be completely specified by the Kraus
operators
\begin{equation}
A_{ij}^{u}=\sqrt{p_{i}p_{j}}\sigma _{i}\otimes \sigma _{j}
\end{equation}%
and with probability $%
%TCIMACRO{\U{3bc} }%
%BeginExpansion
\mu
%EndExpansion
$ the noise is correlated (i.e. the channel has memory) which can be
specified by the Kraus operators
\begin{equation}
A_{kk}^{c}=\sqrt{p_{k}}\sigma _{k}\otimes \sigma _{k}
\end{equation}%
A detailed list of single qubit Kraus operators for different quantum
channels with uncorrelated noise is given in table 1. The action of such a
channel if $n$ qubits are streamed through it, can be described in operator
sum representation as [5]
\begin{equation}
\rho _{f}=\sum\limits_{k_{1,}....,.k_{n}=0}^{n-1}(A_{k_{n}}\otimes
.....A_{k_{1}})\rho _{in}(A_{k_{1}}^{\dagger }\otimes
.....A_{k_{n}}^{\dagger })
\end{equation}%
where $\rho _{in}$ represents the initial density matrix for quantum state
and $A_{k_{n}}$\ are the Kraus operators which satisfy the completeness
relation
\begin{equation}
\sum\limits_{k_{n}=0}^{n-1}A_{k_{n}}^{\dagger }A_{k_{n}}=1
\end{equation}

The Kraus operators for time correlated quantum amplitude damping channel
are given by Yeo and Skeen [40]

\begin{equation}
A_{00}^{c}=\left[
\begin{array}{llll}
\cos \chi & 0 & 0 & 0 \\
0 & 1 & 0 & 0 \\
0 & 0 & 1 & 0 \\
0 & 0 & 0 & 1%
\end{array}%
\right] ,\ \ \ A_{11}^{c}=\left[
\begin{array}{llll}
0 & 0 & 0 & 0 \\
0 & 0 & 0 & 0 \\
0 & 0 & 0 & 0 \\
\sin \chi & 0 & 0 & 0%
\end{array}%
\right]
\end{equation}%
where, $0\leq \chi \leq \pi /2$ and is related to the quantum noise
parameter as
\begin{equation}
\sin \chi =\sqrt{p}
\end{equation}%
The action of such a non-unital channel can be written as
\begin{equation}
\Phi (\rho )=(1-\mu )\sum\limits_{i,j=0}^{1}A_{ij}^{u}\rho A_{ij}^{u\dagger
}+\mu \sum\limits_{k=0}^{1}A_{kk}^{c}\rho A_{kk}^{c\dagger }
\end{equation}

Amplitude damping channel describes how a two-level system approaches the
equilibrium due to coupling with its environment. If the environment has a
finite temperature, then such a dissipation process is described by the
action of a generalized amplitude-damping (GAD) channel with parameter $%
\alpha $ which defines a fixed state of $A_{p,\alpha }$%
\begin{equation}
\rho _{\infty }=\left[
\begin{array}{cc}
\alpha & 0 \\
0 & 1-\alpha%
\end{array}%
\right]
\end{equation}%
If $\alpha =0$ or $\alpha =1,$ the $A_{p,\alpha }$ is simply an amplitude
damping channel. Single qubit Kraus operators for GAD channel are also given
in table 1.

The super-operators provide a neat way of describing the evolution of
quantum states in a noisy environment. In the present scheme, the Kraus
operators are of dimension $2^{4}$. They are constructed from single qubit
Kraus operators by taking their tensor product over all $n^{4}$ combinations
\begin{equation}
A_{k}=\underset{k_{n}}{\otimes }A_{k_{n}}
\end{equation}%
where $n$ is the number of Kraus operator for a single qubit channel. The
final state of the game after the action of the channel can be obtained as
\begin{equation}
\rho _{f}=\Phi _{p,\mu }(\left\vert \Psi \right\rangle \left\langle \Psi
\right\vert )
\end{equation}%
where $\Phi _{p,\mu }$ is the super-operator realizing the quantum channel
parametrized by real numbers $p$ and $\mu $. The players unitary operator is
an $SU(2)$ operator which represents the pure quantum strategy and is given
by
\begin{equation}
\hat{M}(\theta ,\delta ,\beta )=\left[
\begin{array}{cc}
e^{i\delta }\cos (\theta /2) & ie^{i\beta }\sin (\theta /2) \\
ie^{-i\beta }\sin (\theta /2) & e^{-i\delta }\cos (\theta /2)%
\end{array}%
\right]
\end{equation}%
where $0\leq \theta \leq \pi $ and $\pi \leq \{\delta ,$ $\beta \}\leq -\pi
. $ Here $\hat{M}(0,0,0)=\hat{I}$ and $\hat{M}(\pi ,0,0)=i\hat{\sigma}_{x}$
correspond to the two classical pure strategies. Entanglement is controlled
through an entangling gate as given by
\begin{equation}
\hat{J}(\gamma )=\exp (i\frac{\gamma }{2}\sigma _{x}^{\otimes 4})
\end{equation}%
where the parameter $\gamma $ represents the degree of entanglement of the
game and $\gamma =\pi /2$ corresponds to maximal entanglement. It is shown
by Benjamin and Hayden [36] that in a four player quantum Minority game the
Nash equilibrium (NE) strategy is given by
\begin{equation}
\hat{\$}_{NE}=\hat{M}(\frac{\pi }{2},\frac{-\pi }{16},\frac{\pi }{16})
\end{equation}%
This strategic profile $\{\hat{\$}_{NE},$ $\hat{\$}_{NE},$ $\hat{\$}_{NE},$ $%
\hat{\$}_{NE}\}$ results in an NE with an expected payoff of $1/4$. The
expectation value of the payoff to the $k^{\text{th}}$ player can be written
as%
\begin{equation}
\left\langle \$^{k}\right\rangle =\sum\limits_{\xi }\hat{P}_{\xi }\rho _{%
\acute{f}}\hat{P}_{\xi }^{\dagger }\$_{\xi }^{k}
\end{equation}%
where $\hat{P}_{\xi }=\left\vert \xi \right\rangle \left\langle \xi
\right\vert $ is the projector onto the computational state $\left\vert \xi
\right\rangle ,$\ $\$_{\xi }^{k}$ is the payoff to the $k^{\text{th}}$
player when the final state is $\left\vert \xi \right\rangle $ and the
summation is taken over $\xi $ ranging from $1$ to $4$. Let us calculate the
payoff of the first player when all the players resort to their optimal
strategies as given by equation (21). The expected payoff of the first
player influenced by amplitude damping channel reads%
\begin{eqnarray}
\$^{\text{AD}} &=&\frac{1}{8}\mu p^{4}+[\frac{1}{8}(p-1)^{6}+\mu ^{2}\left(
\frac{1}{8}p^{6}-\frac{5}{8}p^{5}+\frac{5}{4}p^{4}-\frac{5}{4}p^{3}+\frac{1}{%
2}p^{2}\right)  \notag \\
&&+\mu \left( -\frac{1}{8}(p-\frac{87}{50})(p-1)^{2}p((p-\frac{44}{25})p+%
\frac{36}{25})\right) ]\sin (\gamma )
\end{eqnarray}%
and the expected payoff of the first player in case of phase flip channel is
given by

\begin{equation}
\$^{\text{PF}}=\frac{1}{8}\left( \left(
\begin{array}{c}
\left.
\begin{array}{c}
-16(\mu -1)^{3}p^{4}+32(\mu -1)^{3}p^{3} \\
-4\left( 5\mu ^{3}-14\mu ^{2}+15\mu -6\right) p^{2}%
\end{array}%
\right. \\
+4\left( \mu ^{3}-2\mu ^{2}+3\mu -2\right) p+1%
\end{array}%
\right) \sin (\gamma )+1\right)
\end{equation}%
The results for the other channels are not presented due to lengthy
relations. The scripts AD, GAD, Dep, BPF and PF in figures correspond to the
amplitude damping, generalized amplitude damping, depolarizing, bit-phase
flip and phase flip channels respectively. Our results are consistent with
Ref. [49] for the case when $\mu $ reaches $0$, as clear from figure 1(c).

\section{Discussions}

The analytical relations for payoffs as a function of decoherence parameter $%
p$, memory parameter $\mu $ and entanglement parameter $\gamma $ are
computed for time correlated amplitude damping, depolarizing, bit-phase flip
and phase flip environments. It is seen that bit-phase flip channel heavily
influences the game as compared to other memory channels.

In figure 1 (a, b), the Nash equilibrium payoff is plotted as a function of
the memory parameter $\mu $ for $\gamma =\pi /2$ and $p=0.3$ and $0.7$ for
amplitude damping, depolarizing, bit-phase flip and phase flip channels
respectively. It is seen that for lower level of decoherence, it is
difficult to distinguish the effect of different channels. However, for
higher level of decoherence, the effect of correlated noise on the dynamics
of the game become more prominent. Therefore, it is quite easy to analyze
the behaviour of different memory channels. In figure 1 (c, d), the Nash
equilibrium payoff is plotted as a function of the decoherence parameter $p$
for $\gamma =\pi /2$ $\mu =0$ and $0.5$ for amplitude damping, depolarizing,
bit-phase flip and phase flip channels respectively. It is seen that the
behaviour of phase flip channel is symmetrical around 50\% decoherence. In
order to see the effect of entanglement on the dynamics of the game in the
presence of decoherence and correlated noise, the Nash equilibrium payoff is
plotted in figure 1 (e) as a function of the entanglement parameter $\gamma $
for $p=\mu =0.3$ for different channels. It is clear from the figure that
amplitude damping channel heavily influences the player's payoff. However,
depolarizing and phase flip channels overlap each other with a similar
behaviour for entire range of entanglement parameter.

In figure 2, the Nash equilibrium payoff is plotted as a function of the
decoherence parameter $p$ and entanglement parameter $\gamma $ for $\mu =0.5$
for amplitude damping, depolarizing, bit-phase flip and phase flip channels
respectively. It is seen that even in the presence of decoherence, the
payoffs are sufficiently enhanced from their classical counterparts due to
the presence of correlated noise. The phase flip channel shows a symmetrical
behaviour around $p=1/2$. Furthermore, the maximum payoff corresponds to the
maximal entanglement situation i.e. $\gamma =\pi /2$ for all the cases. It
is also seen that the role of entanglement is very important during the
course of the game. In figure 3, the Nash equilibrium payoff is plotted as a
function of the decoherence parameter $p$ and memory parameter $\mu $ for $%
\gamma =\pi /2$ for amplitude damping, depolarizing, bit-phase flip and
phase flip channels respectively. It can be inferred from the figure that
the game deserves a careful study during its implementation.

In figure 4, the Nash equilibrium payoff is plotted as a function of (a)
decoherence parameter $p$ and parameter $\alpha $ for $\gamma =\pi /2$ (b)
decoherence parameter $p$ and entanglement parameter $\gamma $ for $\alpha
=1/\sqrt{2}$ (c) parameter $\alpha $ and entanglement parameter $\gamma $
for $p=0.5$ (d) decoherence parameter for different values of parameter $%
\alpha $ for $\gamma =\pi /2$ for the generalized amplitude damping channel.
It is seen that entanglement plays a significant role at lower level of
decoherence (figure 4-b). From figure 4-d, it can be seen that the GAD
channel is less dominant as compared to the AD channel at higher values of
decoherence. However, at lower level of decoherence, both the channels (AD
and GAD) are comparable to each other, where the red curve corresponds to AD
channel, $\alpha =1$ or $0$. Furthermore, the parameter $\alpha $ has
symmetrical effect on the player's payoff (figure 4-c).

\section{Conclusions}

The influence of entanglement and correlated noise on a four-player quantum
Minority game is analyzed. The Nash equilibrium payoff of the first player
is investigated by using different time correlated quantum memory channels.
The players payoffs heavily depends on the memory of the channel. It is seen
that the behaviour of amplitude damping and phase damping channels is
approximately similar. It is shown that bit-phase flip channel heavily
influences the minority game as compared to other channels in the presence
of correlated noise. Whereas, phase flip channel has a symmetrical behaviour
around $50\%$ decoherence. It is seen that entanglement and correlated noise
play a crucial role in minority games. Therefore, the game deserves a
careful study during its implementation. Furthermore, the reduction in
payoffs due to decoherence is well compensated due to the presence of
correlated noise. However, the Nash equilibrium of the game does not change
under correlated noise. Moreover, it is seen that in case of generalized
amplitude damping channel, entanglement plays a significant role at lower
level of decoherence. It has smaller damping effects on the payoff at higher
values of decoherence $(p>0.5)$ as compared to the amplitude damping
channel. However, amplitude damping and generalized amplitude damping
channels have similar effect on the player's payoffs at lower level of
decoherence $(p<0.5)$.\newline

\begin{figure}[tbp]
\begin{center}
\vspace{-2cm} \includegraphics[scale=0.8]{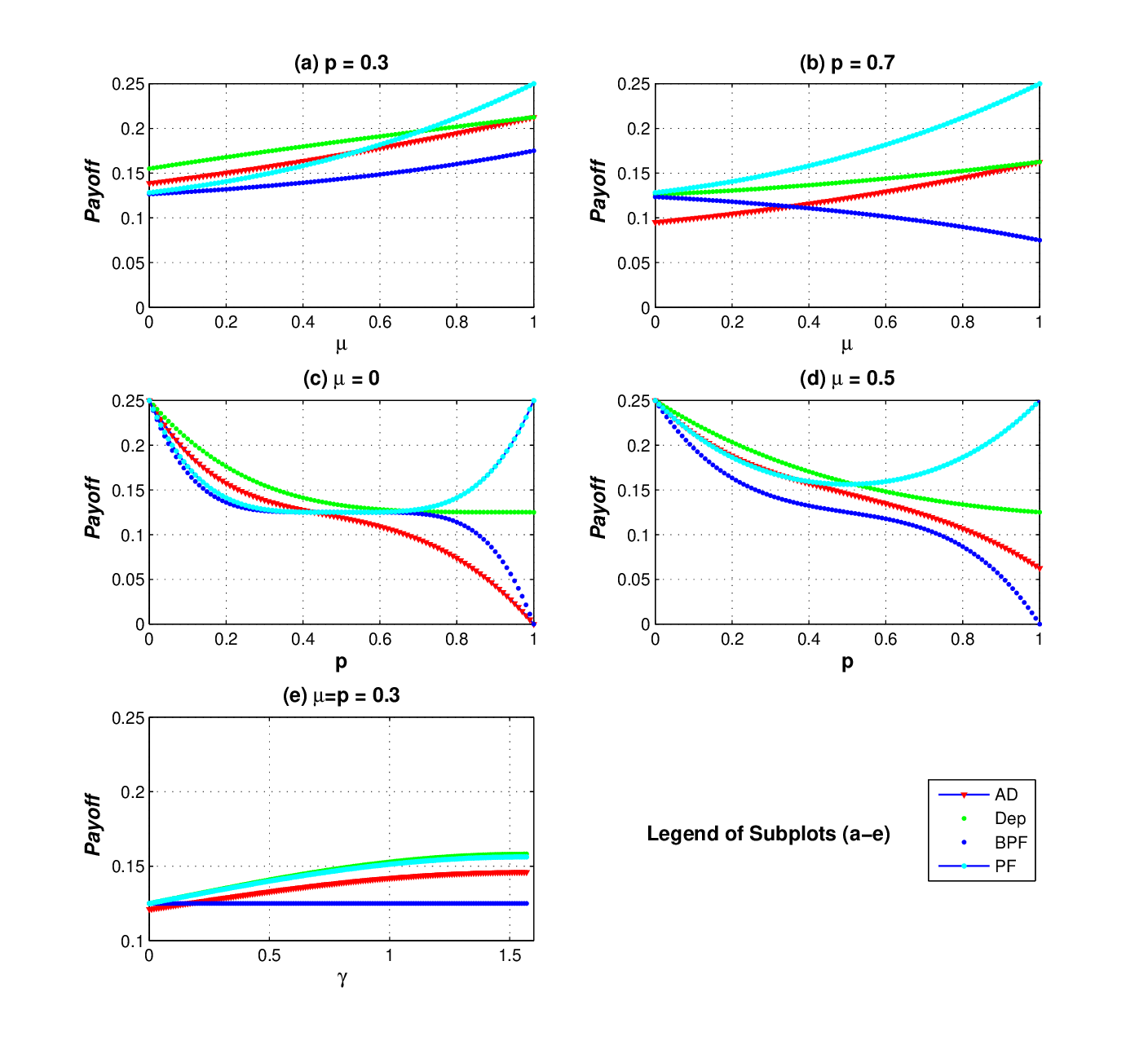} \\[0pt]
\end{center}
\caption{(Color online). The Nash equilibrium payoff is plotted as a
function of the memory parameter $\protect\mu $ for $\protect\gamma =\protect%
\pi /2$ and (a, b) $p=0.3$ and $0.7$, as a function of the decoherence
parameter $p$ for $\protect\gamma =\protect\pi /2$ and (c, d) $\protect\mu %
=0 $ and $0.5$, as a function of the entanglement parameter $\protect\gamma $
(e) for $p=\protect\mu =0.3$ for amplitude damping, depolarizing, bit-phase
flip and phase flip channels respectively.\newline
}
\end{figure}

\begin{figure}[tbp]
\begin{center}
\vspace{-2cm} \includegraphics[scale=0.8]{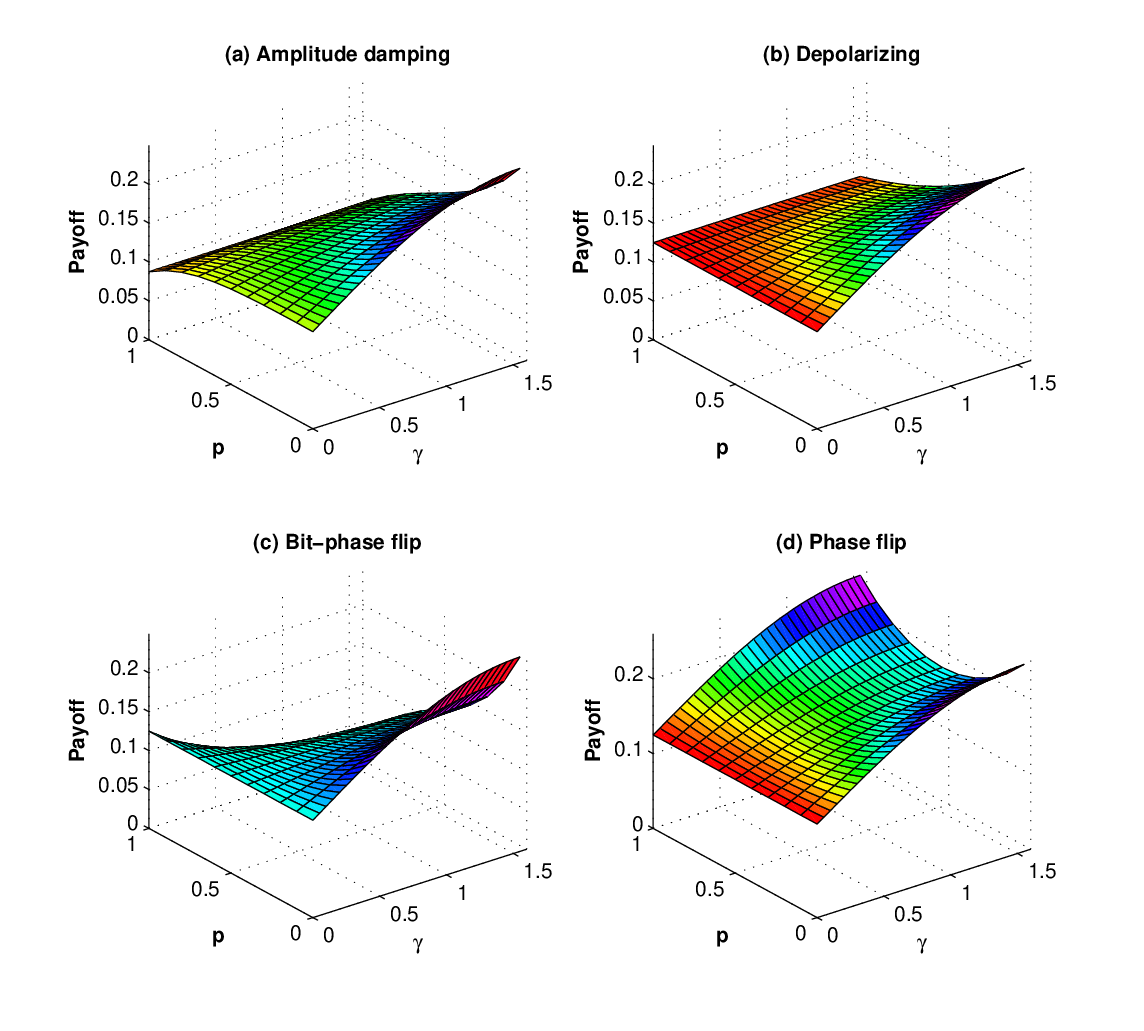} \\[0pt]
\end{center}
\caption{(Color online). The Nash equilibrium payoff is plotted as a
function of the decoherence parameter $p$ and entanglement parameter $%
\protect\gamma $ for $\protect\mu =0.5$ for amplitude damping, depolarizing,
bit-phase flip and phase flip channels respectively.\newline
}
\end{figure}

\begin{figure}[tbp]
\begin{center}
\vspace{-2cm} \includegraphics[scale=0.8]{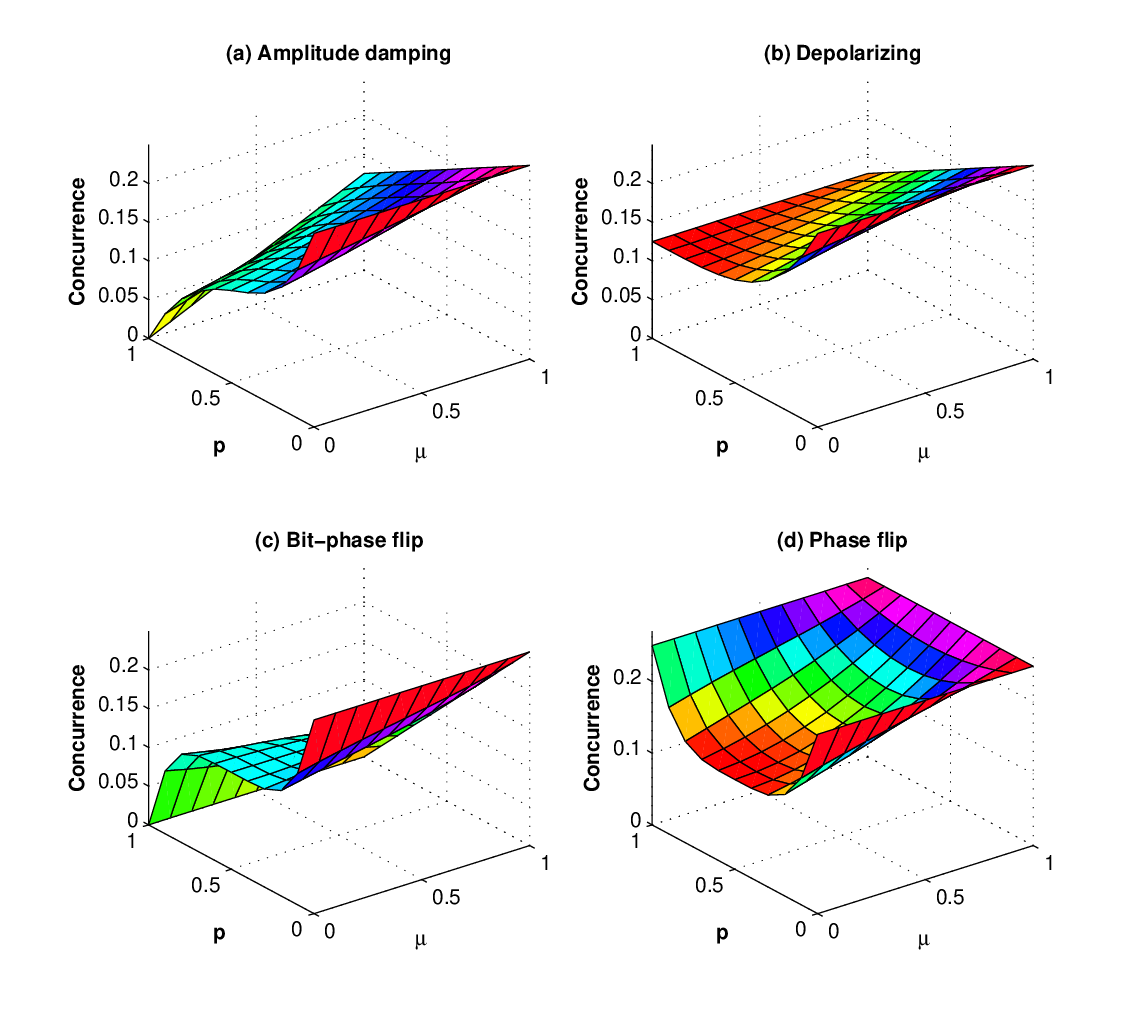} \\[0pt]
\end{center}
\caption{ (Color online). The Nash equilibrium payoff is plotted as a
function of the decoherence parameter $p$ and memory parameter $\protect\mu $
for $\protect\gamma =\protect\pi /2$ for amplitude damping, depolarizing,
bit-phase flip and phase flip channels respectively.\newline
}
\end{figure}

\begin{figure}[tbp]
\begin{center}
\vspace{-2cm} \includegraphics[scale=0.8]{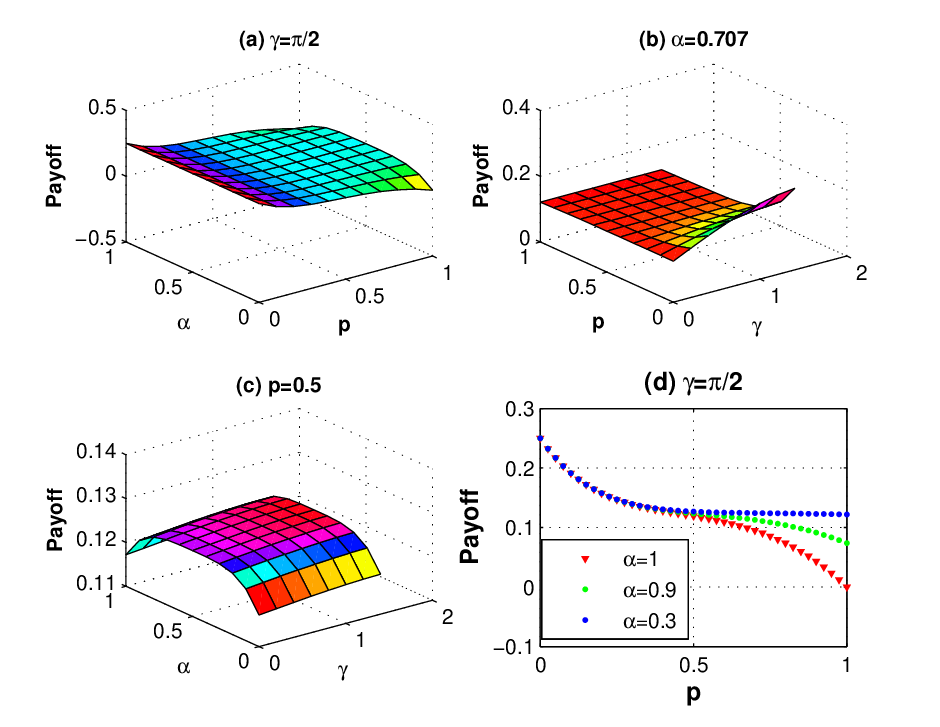} \\[0pt]
\end{center}
\caption{ (Color online). The Nash equilibrium payoff is plotted as a
function of the decoherence parameter $p$, parameters $\protect\alpha $ and $%
\protect\gamma $ for different situations as labeled on subgraphs (a-d) for
generalized amplitude damping channel.\newline
}
\end{figure}

\begin{table}[tbh]
\caption{Single qubit Kraus operators for amplitude damping, generalized
amplitude damping, depolarizing, bit-phase flip, bit flip and phase flip
channels where $p$ represents the decoherence parameter and parameter }$%
\begin{tabular}{|l|l|}
\hline
&  \\
$\text{Amplitude damping channel}$ & $A_{0}=\left[
\begin{array}{cc}
1 & 0 \\
0 & \sqrt{1-p}%
\end{array}%
\right] ,$ $A_{1}=\left[
\begin{array}{cc}
0 & \sqrt{p} \\
0 & 0%
\end{array}%
\right] $ \\ \hline
$%
\begin{tabular}{l}
Generalized a$\text{mplitude}$ \\
$\text{damping channel}$%
\end{tabular}%
$ & $%
\begin{tabular}{l}
$A_{0}=\sqrt{\alpha }\left[
\begin{array}{cc}
1 & 0 \\
0 & \sqrt{1-p}%
\end{array}%
\right] ,\quad A_{1}=\sqrt{\alpha }\left[
\begin{array}{cc}
0 & \sqrt{p} \\
0 & 0%
\end{array}%
\right] $ \\
$A_{2}=\sqrt{1-\alpha }\left[
\begin{array}{cc}
\sqrt{1-p} & 0 \\
0 & 0%
\end{array}%
\right] ,A_{3}=\sqrt{1-\alpha }\left[
\begin{array}{cc}
0 & 0 \\
\sqrt{p} & 0%
\end{array}%
\right] $%
\end{tabular}%
$ \\ \hline
$\text{Depolarizing channel}$ & $%
\begin{tabular}{l}
$A_{0}=\sqrt{1-\frac{3p}{4}I},\quad A_{1}=\sqrt{\frac{p}{4}}\sigma _{x}$ \\
$A_{2}=\sqrt{\frac{p}{4}}\sigma _{y},\quad \quad $\ $\ A_{3}=\sqrt{\frac{p}{4%
}}\sigma _{z}$%
\end{tabular}%
$ \\
&  \\ \hline
$\text{Bit-phase flip channel}$ & $A_{0}=\sqrt{1-p}I,\quad A_{1}=\sqrt{p}%
\sigma _{y}$ \\
&  \\ \hline
$\text{Bit flip channel}$ & $A_{0}=\sqrt{1-p}I,\quad A_{1}=\sqrt{p}\sigma
_{x}$ \\
&  \\ \hline
$\text{Phase flip channel}$ & $A_{0}=\sqrt{1-p}I,\quad A_{1}=\sqrt{p}\sigma
_{z}$ \\ \hline
\end{tabular}%
$%
\label{di-fit}
\end{table}

\end{document}